\documentclass[12pt]{article}
\usepackage{amsfonts,amssymb}
\begin{document}
\vspace*{1cm}
\begin{center}
{\Large \bf Hadronic Spectra and \\[1ex] Kaluza-Klein Picture of the World}
\footnote{Extended version of the talk presented at Xth International
Conference on Hadron Spectroscopy HADRON '03 , August~31 --
September~6, 2003, Aschaffenburg, Germany.}

\vspace{4mm}

{\large A.A. Arkhipov\\
{\it State Research Center ``Institute for High Energy Physics" \\
 142280 Protvino, Moscow Region, Russia}}\\
\end{center}

\vspace{4mm}
\begin{abstract}
{A manifestation of Kaluza-Klein picture in hadronic spectra is
discussed. We argue that the experimentally observed structures in
hadronic spectra confirm the Kaluza-Klein picture of the world.}
\end{abstract}

\vspace{4mm}

\rightline{\it\large ``... the simpler the presentation of a
particular law of Nature,}

\vspace{2mm}

\rightline{\it\large the more general it is ..."}

\vspace{4mm}

\rightline{Max Planck, Nobel Lecture, June 2, 1920}

\vspace{4mm}

\section{Introduction}

\noindent Dear Colleagues.

It seems that here is just the place where we could remember one of
the greatest physicists of the  XXth century, I mean German physicist
Max Planck. My experience in science allows me to definitely share
Max Planck's opinion in the above written fragment of his Nobel
Lecture. Following this opinion, I'd like to present here a new, very
simple and at the same time quite general, physical law concerning
the structure of hadron spectra.

Although the modern strong interaction theory formulated in terms of
the known QCD Lagrangian is commonly accepted, this theory does not
allow
 to make an appreciable breakthrough in the problem of calculating
the masses of compound systems so far, mainly because that problem is
significantly non-perturbative. In other words, this means that our
theoretical understanding of low-energy QCD spectroscopy is far from
what is desired. Even the best currently performed lattice
computations in QCD cannot help us to understand the exact nature of
the real hadron spectrum.

All of you know that strong interactions are characterized by
multi-particle production. The dynamics of multi-particle systems
necessarily involve so called many-body forces. Many-body forces are
fundamental forces arising in multi-particle systems with more than
two particles, and they are responsible for the dynamics of
production processes. For example, the three-body forces are
responsible for the dynamics of one-particle inclusive reactions; see
Ref. \cite{1} and references therein. A description of many-body
forces requires the use of multidimensional spaces, and, as a
consequence, the strong interactions theory cannot be constructed
consistently if multidimensional spaces are not used. Therefore, it
seems natural to formulate the strong interactions theory in a
multidimensional space from the very beginning.

The idea to use multidimensional spaces in fundamental physics is not
new: famous works by Kaluza and Klein were the first to introduce
this idea. The original idea of Kaluza and Klein is based on the
hypothesis that the input space-time is a $(4+d)$-dimensional space
${\cal M}_{(4+d)}$ which can be represented as a tensor product of
the visible four-dimensional world $M_4$ with a compact internal
$d$-dimensional space ${\cal K}_d$
\begin{equation}
{\cal M}_{(4+d)} = M_4 \times {\cal K}_d.\label{1}
\end{equation}
The compact internal space ${\cal K}_d$ is space-like, i.e. it has
only spatial dimensions which may be considered as extra spatial
dimensions of $M_4$. An especial example of ${\cal M}_{(4+d)}$ is a
space with a factorizable metric. According to the tensor product
structure of the space ${\cal M}_{(4+d)}$, the metric may be chosen
in a factorizable form. This means that if $z^M = \{ x^{\mu},
y^{m}\}$, ($M=0,1,\ldots,3+d,\, \mu = 0,1,2,3,\, m=1,2, \ldots, d$)
are local coordinates on ${\cal M}_{(4+d)}$, then the factorizable
metric looks like
\[
ds^{2} = {\cal G}_{MN}(z) dz^M dz^N = g_{\mu \nu}(x) dx^{\mu}
dx^{\nu} + \gamma_{mn}(x,y) dy^{m} dy^{n},
\]
where $g_{\mu \nu}(x)$ is the metric on $M_4$.

In the year 1921, Kaluza proposed a unification of the theory of
gravity and Maxwell theory of electromagnetism in four dimensions
starting from the theory of gravity in five dimensions. He assumed
that the five-dimensional space ${\cal M}_5$ had to be the product of
the four-dimensional space-time $M_4$ and a circle ${\cal S}_1$:
${\cal M}_5 = M_4 \times{\cal S}_1$. It was shown that the zero mode
sector of the Kaluza model is equivalent to the four-dimensional
theory which describes the Hilbert-Einstein gravity with
four-dimensional general coordinate transformations and the Maxwell
theory of electromagnetism with gauge transformations.

Recently, some models with extra dimensions have been proposed to
attack the electroweak quantum instability of the Standard Model
known as the hierarchy problem between electroweak and gravity
scales. However, it is obvious that the basic idea of the
Kaluza-Klein scenario may be applied to any model in the Quantum
Field Theory. As an illustrative example, let us consider the
simplest case of the (4+d)-dimensional model of scalar field with the
action
\begin{equation}
S = \int d^{4+d}z \sqrt{-{\cal G}} \left[ \frac{1}{2} \left(
\partial_{M} \Phi \right)^2 - \frac{m^{2}}{2} \Phi^2 +
\frac{G_{(4+d)}}{4!} \Phi^4 \right], \label{S}
\end{equation}
where ${\cal G}=\det|{\cal G}_{MN}|$, ${\cal G}_{MN}$ is the metric
on ${\cal M}_{(4+d)} = M_4 \times{\cal K}_d$, $M_4$ is the
pseudo-Euclidean Minkowski space-time, ${\cal K}_d$ is a compact
internal $d$-dimensional space with the characteristic size $R$. Let
$\Delta_{{\cal K}_d}$ be the Laplace operator on the internal space
${\cal K}_d$, and $Y_{n}(y)$ ortho-normalized eigenfunctions of the
Laplace operator
\begin{equation}
\Delta_{{\cal K}_{d}} Y_{n}(y) = -\frac{\lambda_{n}}{R^{2}} Y_{n}(y),
\label{Yn}
\end{equation}
here $n$ is the (multi)index labelling the eigenvalue $\lambda_{n}$
of the eigenfunction $Y_{n}(y)$. A $d$-dimen\-sional torus ${\cal
T}^{d}$ with equal radii $R$ is an especially simple example of the
compact internal space of extra dimensions ${\cal K}_d$. The
eigenfunctions and eigenvalues in this special case look like
\begin{equation}
Y_n(y) = \frac{1}{\sqrt{V_d}} \exp \left(i \sum_{m=1}^{d}
n_{m}y^{m}/R \right), \label{T}
\end{equation}
\[
\lambda_n = |n|^2,\quad |n|^2= n_1^2 + n_2^2 + \ldots n_d^2, \quad
n=(n_1,n_2, \ldots, n_d),\quad -\infty \leq n_m \leq \infty,
\]
where $n_m$ are integer numbers, $V_d = (2\pi R)^d$ is the volume of
the torus.

To reduce the multidimensional theory to the effective
four-dimensional one we write a harmonic expansion for the
multidimensional field $\Phi(z)$
\begin{equation}
\Phi(z) = \Phi(x,y) = \sum_{n} \phi^{(n)}(x) Y_{n}(y). \label{H}
\end{equation}
The coefficients $\phi^{(n)}(x)$ of harmonic expansion (\ref{H}) are
called Kaluza-Klein (KK) excitations or KK modes, and they usually
include the zero-mode $\phi^{(0)}(x)$, corresponding to $n=0$ and the
eigenvalue $\lambda_{0} = 0$. Substitution of the KK mode expansion
into action (\ref{S}) and integration over the internal space ${\cal
K}_{d}$ gives
\begin{equation}
S = \int d^{4}x \sqrt{-g} \left\{ \frac{1}{2} \left(
\partial_{\mu} \phi^{(0)} \right)^{2} - \frac{m^{2}}{2}
(\phi^{(0)})^{2} \right. + \frac{g}{4!} (\phi^{(0)})^{4} +
\end{equation}
\[
+\left. \sum_{n \neq 0} \left[\frac{1}{2} \left(\partial_{\mu}
\phi^{(n)} \right) \left(\partial^{\mu} \phi^{(n)} \right)^{*} -\frac
{m_n^2}{2} \phi^{(n)}\phi^{(n)*} \right] + \frac{g}{4!}
(\phi^{(0)})^{2} \sum_{n\neq 0} \phi^{(n)} \phi^{(n)*}\right\} +
\ldots.
\]
For the masses of the KK modes one obtains
\begin{equation}\label{m}
\fbox{$\displaystyle m_{n}^{2} = m^{2} + \frac{\lambda_{n}}{R^2}$}\,,
\end{equation}
and the coupling constant $g$ of the four-dimensional theory is
related to the coupling constant $G_{(4+d)}$ of the initial
multidimensional theory by the equation
\begin{equation}
\fbox{$\displaystyle  g = \frac{G_{(4+d)}}{V_d}$}\,,\label{g}
\end{equation}
where $V_d$ is the volume of the compact internal space of extra
dimensions ${\cal K}_d$. The fundamental coupling constant
$G_{(4+d)}$ has dimension $[\mbox{mass}]^{-d}$. So, the
four-dimensional coupling constant $g$ is dimensionless, as it should
be. Eqs.~(\ref{m},\ref{g}) represent the basic relations of the
Kaluza-Klein scenario.  Similar relations take place for other types
of multidimensional quantum field theoretical models. From the
four-dimensional point of view we can interpret each KK mode as a
particle with the mass $m_n$ given by Eq.~(\ref{m}). Clearly,
according to the Kaluza-Klein scenario any multidimensional field
contains an infinite set of KK modes, i.e. an infinite set of
four-dimensional particles with increasing masses, which is called
the Kaluza-Klein tower. Therefore, an experimental observation of the
series of KK excitations with a characteristic spectrum of form
(\ref{m}) would be an evidence of the existence of extra dimensions.
So far, the KK partners of the particles of the Standard Model have
not been observed. In the Kaluza-Klein scenario this fact can be
explained by a microscopically small size $R$ of extra dimensions
($R<10^{-17}\,cm$); in that case the KK excitations may be produced
only at super-high energies of the scale $E\sim 1/R > 1\,TeV$. Below
this scale, only homogeneous zero modes with $n=0$ were accessible
for observation in recent high energy experiments. That is why, there
is a hope to search the KK excitations at the future LHC and other
colliders.

As we have calculated before \cite{2}
\begin{equation}\label{sc}
\frac{1}{R}=41.481 \mbox{MeV},
\end{equation}
or
\begin{equation}\label{size}
\fbox{$\displaystyle R=24.1\,GeV^{-1}=4.75\,10^{-13}\mbox{cm}$}\,.
\end{equation}
If we relate the strong interaction scale with the pion mass
\begin{equation}\label{G}
G_{(4+d)}\sim\frac{10}{[m_\pi]^d},
\end{equation}
then
\begin{equation}\label{simg}
g\sim\frac{10}{(2\pi m_\pi R)^d},
\end{equation}
and
\[
g(d=1)\sim 0.5.
\]
On the other hand
\begin{equation}\label{geff}
g_{eff}=g_{\pi NN}\exp(-m_{\pi}R)\sim 0.5,\ \ \  (g^2_{\pi
NN}/4\pi=14.6).
\end{equation}
So, $R$ has a clear physical meaning: size (\ref{size}) just
corresponds to the scale of distances where strong Yukawa forces in
strength come close to electromagnetic ones. Moreover,
\begin{equation}\label{SM}
M\sim R^{-1}\left(M_{Pl}/R^{-1}\right)^{2/(d+2)}\mid_{d=6}\, \sim
5\,\mbox{TeV}.
\end{equation}
Mass scale (\ref{SM}) is just the scale accepted in the Standard
Model, and this is an interesting observation as well. Actually, mass
scale (\ref{SM}) means that gravity effects may be detected at the
future LHC collider.

\section{Peculiarities of Kaluza-Klein excitations}

From the formula for the masses of the KK modes
\[
 m_{n} = \sqrt{m^{2} + \frac{n^2}{R^2}}
\]\\
we obtain
\begin{equation}\label{potbox}
m_n = m + \delta m_n,\quad \delta m_n = \frac{n^2}{2mR^2},\quad
n<<mR,
\end{equation}
and this just corresponds to the spectrum of the potential box with
the size equal to the size of the internal compact extra space. In
the other case,
\begin{equation}\label{Coulomb}
m_n = n\omega + \delta m_n,\quad \delta m_n =
\frac{m\alpha^2}{2n},\quad \omega \equiv \frac{1}{R},\quad \alpha^2
\equiv mR,\quad \alpha^2<<n,
\end{equation}
and here we come to the (quasi)oscillator (quasi, because $n$ takes
place of $n+1/2$ for the one-dimensional case) and (quasi)Coulomb
(with  $1/n$ instead of $1/n^2$ and $\alpha^2=mR$ instead of
$\alpha_c^2=(1/137)^2$) spectra. Clearly, we can neglect the
(quasi)Coulomb contribution in the region $n>>\alpha^2\equiv mR$.

It is remarkable that KK modes of relativistic origin due to
quantization of finite motion in the space of extra dimensions
interpolate the non-relativistic spectrum of the potential box and
the oscillator spectrum.

The spectrum of the two($a$ and $b$)-particle compound system is
defined in the fundamental (input) theory by the formula
\begin{equation}\label{comp}
M_{ab}^n = m_a + m_b + \delta m_{ab}^n(m_a,m_b,G_{4+d}).
\end{equation}
The goal of the fundamental theory is to calculate $\delta
m_{ab}^n(m_a,m_b,G_{4+d})$. There is no solution of that problem in
the strong interaction theory because this is a significantly
non-perturbative problem. However, in the framework of Kaluza-Klein
approach we can rewrite the above formula in an equivalent form
\begin{equation}\label{KKcomp}
M_{ab}^n = m_{a,n} + m_{b,n} + \delta m_{ab,n}(m_{a,n},m_{b,n},g),
\end{equation}
where $m_{a,n},m_{b,n}$ are KK modes of particles $a$ and $b$, and
the quantity $\delta m_{ab,n}(m_{a,n},m_{b,n},g)$  can be calculated
using the four-dimensional perturbation theory. Moreover, because
$\delta m_{ab,n}(m_{a,n},m_{b,n},g)<<m_{a(b),n}$, we can put with a
high accuracy
\begin{equation}\label{approx}
M_{ab}^n \cong m_{a,n} + m_{b,n},
\end{equation}
and this allows to formulate the global solution of the spectral
problem in hadron spectroscopy.

\section{On the global solution of the spectral problem}

According to Kaluza and Klein,  we suggest that the input
(fundamental) space-time ${\cal M}_{(4+d)}$ is represented as
\[
{\cal M}_{(4+d)} = M_4 \times {\cal K}_d.
\]
Let $\lambda_{n}$ be characteristic numbers of the Laplace operator
on ${\cal K}_{d}$ with a characteristic size $R_{\cal K}$
\[
\Delta_{{\mathcal K}_{d}} Y_{n}(y) = -\frac{\lambda_{n}}{R_{\cal
K}^{2}} Y_{n}(y).
\]\\
Let $\lambda_{\cal K}$ be the set of all characteristic numbers of
the Laplace operator
\begin{equation}\label{charset}
\lambda_{\cal K} \equiv \biggl\{ \lambda_n: n\in {\mathbb Z}^d \equiv
\underbrace{{\mathbb Z}\times{\mathbb Z}\times\cdots\times{\mathbb
Z}}_d\, \biggr\}.
\end{equation}
There is one-to-one correspondence
\[
{\cal K} \quad \Longleftrightarrow \quad (R_{\cal K},\lambda_{\cal
K}).
\]
Let us consider a compound hadron system $h$ which may decay into
some channel
\begin{equation}\label{channel}
h \rightarrow a+b+\cdots +c.
\end{equation}
We introduce the spectral mass function of the given channel by the
formula
\begin{equation}\label{spmass}
M_h^{ab...c}(R_{\cal
K},\lambda_{n_a},\lambda_{n_b},\cdots,\lambda_{n_c})=\sqrt{m_a^2 +
\frac{\lambda_{n_a}}{R_{\cal K}^2}} + \sqrt{m_b^2 +
\frac{\lambda_{n_b}}{R_{\cal K}^2}} + \cdots + \sqrt{m_c^2 +
\frac{\lambda_{n_c}}{R_{\cal K}^2}}.
\end{equation}
Now we build the Kaluza-Klein tower:
\begin{equation}\label{tower}
t_h^{ab...c}({\cal K})\equiv t_h^{ab...c}(R_{\cal K},\lambda_{\cal
K})\stackrel{def}{=} \biggl\{M_h^{ab...c}(R_{\cal
K},\lambda_{n_a},\lambda_{n_b},\cdots,\lambda_{n_c}):
\lambda_{n_i}\in\lambda_{\cal K}\, (i=a,b,...,c) \biggr\}.
\end{equation}
After that we build the Kaluza-Klein town as a union of the
Kaluza-Klein towers corresponding to all possible decay channels of
the hadron system $h$
\begin{equation}\label{town}
{\cal T}_h({\cal K}) \equiv {\cal T}_h(R_{\cal K},\lambda_{\cal
K})\stackrel{def}{=} \bigcup_{\{ab...c\}}t_h^{ab...c}(R_{\cal
K},\lambda_{\cal K}).
\end{equation}

We state:
\begin{equation}\label{hmass}
\fbox{$\displaystyle M_h \in {\cal T}_h({\cal K})$}\,.
\end{equation}
Let ${\cal H}$ be the set of all possible physical hadron states. We
build the hadron Kaluza-Klein country ${\mathbb C}_{\cal H}({\cal
K})$ by the formula
\begin{equation}\label{country}
{\mathbb C}_{\cal H}({\cal K}) \stackrel{def}{=} \bigcup_{h\in {\cal
H}}{\cal T}_h({\cal K}).
\end{equation}
The whole spectrum of all possible physical hadron states we denote
as $M_{\cal H}$
\begin{equation}\label{spectr}
M_{\cal H} \stackrel{def}{=} \biggl\{M_h: h\in {\cal H}\biggr\}.
\end{equation}

We state:
\begin{equation}\label{globsol}
\fbox{$\displaystyle M_{\cal H} \in {\mathbb C}_{\cal H}({\cal
K})$}\,.
\end{equation}
Formulae (\ref{hmass}) and (\ref{globsol}) provide the global
solution of the spectral problem in hadron spectroscopy.

Here is just the place to make some clarifying remarks. First of all,
in the construction of the global solution among all possible decay
channels of the hadron system $h$ only those channels should be taken
into account which contain fundamental particles and their different
multi-particle compound systems in the final states, as it really
should be. Appearance of non-zero KK modes of the fundamental
particles and their compound systems in the final states of the decay
channels is forbidden by the construction. For example, the decay
channel
\begin{equation}\label{channel2}
h \nrightarrow a^*+b+\cdots +c,
\end{equation}
where $a^*$ is a non-zero KK mode of the fundamental particle $a$,
cannot be used in the construction. The decay channel
\begin{equation}\label{channel3}
h \rightarrow A+b+\cdots +c,
\end{equation}
where $A$ is some multi-particle compound system which may decay into
some channel with the fundamental particles $a_i (i=1,2,...k)$ in the
final state
\begin{equation}\label{channel4}
A \rightarrow a_1+a_2+\cdots +a_k,
\end{equation}
is admissible by the construction. But the decay channel
\begin{equation}\label{channel5}
h \nrightarrow A^*+b+\cdots +c,
\end{equation}
where $A^*$ denotes some non-zero KK mode of $A$, is forbidden. In
other words, the underlying physical principle in the construction of
the global solution was the principle of non-observability of
non-zero KK modes of the fundamental particles and their compound
systems. According to that principle, non-zero KK modes of the
fundamental particles may manifest themselves only virtually during
an interaction, for example while they are staying in a compound
system. Non-zero KK modes of the fundamental particles living in a
compound system define the main properties of a compound system, such
as mass and life time of a system. As was mentioned above, an
interaction of KK modes is weak, therefore we can calculate with a
high accuracy the mass of a compound system as a simple sum of the
masses of KK modes. Moreover, weakly interacting KK modes result in
very narrow widths of the compound states, and this phenomenon is
observed in recent experiments.

The dynamics of the decays of compound systems is physically
transparent: non-zero KK modes of the constituents make a transition
to zero KK modes, and we observe zero KK modes as decay products. In
the framework of such decay dynamics, we can estimate the widths of
the compound states
\begin{equation}\label{width}
\Gamma_n \sim \frac{\alpha_{eff}}{2}\cdot\frac{n}{R}\cdot O(1)\sim
0.4\cdot n \, \mbox{MeV},
\end{equation}
where $n$ is the KK excitation number. The broad peaks in the hadron
spectra are interpreted as an envelope of the narrow peaks predicted
by the Kaluza-Klein scenario.

We have shown in the previous section that non-zero KK modes look
like the states of a particle in confining potentials. Such a
particle might be considered as a quasi-particle which cannot be
observed without destroying a confining potential. The quasi-particle
becomes a real particle by the transition of a non-zero KK mode to a
zero KK mode which is equivalent to destroying the confining
potential, and we observe a zero KK mode i.e. a real fundamental
particle as a decay product. This consideration justifies the
underlying physical principle in the construction of the global
solution. In fact, here quite a new approach to the Kaluza-Klein
picture as a whole is presented.

\section{Comparison with the experimental data}

In Refs. \cite{2,3,4,6,8,12} we verified the global solution with the
set of experimental data for the two-nucleon system, two-pion system,
three-pion system, strange mesons, charmed and charmed-strange
mesons, and found out that the solution described accurately the
experimentally observed hadron spectra. Here we extract the main
Tables from the previous papers with an addition of some new ones.

Table 1 extracted from Ref. \cite{3} contains the theoretically
calculated Kaluza-Klein tower of KK excitations for the two-nucleon
system with the account that Kaluza-Klein scenario predicts
$M_n^{pp}=M_n^{p\bar p}$, and experimentally observed mass spectra of
proton-proton and proton-antiproton systems above the elastic
threshold. The sources of experimental data see in literature of Ref.
\cite{3}. As is seen from Table 1, the nucleon-nucleon dynamics at
low energies provides quite a remarkable confirmation of Kaluza-Klein
picture. Moreover, Kaluza-Klein scenario predicts a special sort of
(super)symmetry between fermionic (dibaryon) and bosonic states,
which is quite nontrivial, and Table 1 contains an experimental
confirmation of this fact as well. We do hope that blanks in Table 1
will be filled in the future experimental studies.

The Kaluza-Klein tower of KK excitations for the two-pion system,
extracted from Ref.~\cite{4}, is shown in Table 2 where the
comparison with experimentally observed mass spectrum of two-pion
system is also presented; see details in \cite{4}. Here we have only
one empty cell $M_{13}^{\pi\pi}(1112-1114)$, and this is the subject
for  further careful analysis of the two-pion system. Besides, we
would like to draw your attention to reference \cite{5}\footnote{I
thank E.~Kolomeitsev for drawing my attention to this article.},
which contains a review of the known earlier results concerning the
ABC-particle observed for the first time in 1961 at Berkeley in the
reaction $pd\rightarrow {}^3HeX^0$. A more precise experimental study
in Ref.~\cite{5} compared to the experiments performed before allowed
to establish four states in the two-pion system with the masses
$M\approx 310$ MeV, $M\approx 350$ MeV, $M\approx 430$ MeV, $M\approx
550$ MeV. From expressive, historical report \cite{6} presented at
this Conference I have learned about old work \cite{7} where the
two-pion effective-mass distributions in the reaction of $\bar pp$
annihilation have been studied in 1962 at Berkeley too. It should be
emphasized that this important paper was not cited in Ref.~\cite{5}.
However, I found this article quite an interesting. First of all, as
it follows from Ref. \cite{7}, measuring $K_1^0$ decays produced by
$\bar p+p\rightarrow K_1^0+K+n\pi$, a peak in the  two-pion
effective-mass distribution with $\Gamma/2=13$ MeV centered at
$499\pm 3$ MeV has been observed. This just corresponds to the
$M_5^{\pi\pi}$-storey in Table 2. Secondly, a careful inspection of
the 750-MeV region in the two-pion effective-mass distributions
revealed an evidence for a double peak in this region. It was
concluded that the data satisfy very well the hypothesis of two peaks
at 720 and 780 MeV with $\Gamma_1=30$ MeV and  $\Gamma_2=60$ MeV: the
data are best fit by these two peaks. We just predict these states;
see $M_8^{\pi\pi}$-storey and $M_9^{\pi\pi}$-storey in Table 2.
Finally, a remarkable discontinuity in  the two-pion effective-mass
spectrum at 320 MeV has been observed \cite{7} which referred to ABC
peak. $M_2^{\pi\pi}$-storey in Table 2 just corresponds to this peak.
Thus,  Table 2 shows that there is quite a remarkable correspondence
of the calculated KK excitations for the two-pion system with the
experimentally observed picture in the two-pion effective-mass
spectrum, and this can be considered as strong evidence of
Kaluza-Klein picture of the world. It should additionally be pointed
out that the $f_2(0^+2^{++})$-mesons ($M_{f_2} = 1272\pm 8$ MeV  and
$M_{f_2} = 2175\pm 20$ MeV) investigated by IHEP group accurately
agree with the calculated values and excellently incorporated in the
scheme of systematics provided by Kaluza-Klein picture \cite{4}.

The Kaluza-Klein tower built for the three-pion system in Ref.
\cite{8} is shown in Table 3 where the comparison with experimentally
observed mass spectrum of the three-pion system is also presented.
Certainly, here is a much more poor experimental data set compared to
the case of the two-pion system. Nevertheless, again we see from
Table 3 that there is quite a remarkable correspondence of the
calculated KK excitations for the three-pion system with the
experimental data where such data exist, and this fact can also be
considered  as an additional evidence of Kaluza-Klein picture of the
world. It is pleased for us to emphasize \cite{8} that the
experimental measurement of the $a_2$ meson mass made by Protvino VES
Collaboration with the best world precision
\[
M(a_2)=1311.3 \pm 1.6(stat) \pm 3.0(syst)\,\mbox{MeV}
\]
is in excellent agreement with the theoretically calculated value
\[
M_{10}^{\pi^+\pi^-\pi^0}=1311.55\,\mbox{MeV}.
\]
The same is true for the $\omega_3$ meson where the theoretically
calculated mass of KK excitation in the  $3\pi^0$ system
$M_{13}^{3\pi^0}=1667.68\,MeV$ is in a very good agreement with PDG
AVERAGE value $M(\omega_3)=1667 \pm 4\,MeV$. Moreover, it is very
interesting to point out that theoretical calculation of KK
excitations in the $\rho\pi$ system by the formula
\begin{equation}
M_n^{\rho\pi} = \sqrt{m_{\rho}^2+\frac{n^2}{R^2}} +
\sqrt{m_{\pi}^2+\frac{n^2}{R^2}},\quad (n=1,2,3,\ldots),\label{rhopi}
\end{equation}
where we use $m_{\rho}=769.3\,MeV$ for the $\rho$ meson mass from
PDG, gives $M_{10}^{\rho\pi^0}=1310.28\,MeV$ and
$M_{10}^{\rho\pi^{\pm}}=1311.67\,MeV$ which accurately agree with the
experimental measurement of the $a_2$ meson mass provided by VES
Collaboration. This means that $a_2$ meson may manifest itself as a
configuration of the $\rho\pi$ system in the main, and this is quite
nontrivial. For example, that statement is not true for the
$\omega_3$ meson. The Kaluza-Klein tower built for the
$\rho\pi$-system is shown in Table 4. VES Collaboration gives
$M(\pi_1(1^{-+}))=1610\pm20\,MeV, (\Gamma=290\pm30\,MeV$, see
\cite{9}), which is excellently incorporated in Table 4.

Table 5 corresponds to the Kaluza-Klein tower of KK excitations for
the two-kaon system built in Ref. \cite{10}. This Table apart of old
experimental data contains new states recently observed in
$K_S^0K_S^0$-system and presented at this Conference in the talks
\cite{11,12}. It is very nice to emphasize that all new states are in
excellent agreement with the values predicted in the scheme of the
systematics provided by Kaluza-Klein approach.

Table 6 extracted from Ref. \cite{10} shows the Kaluza-Klein tower of
KK excitations for the $K\pi$-system, and experimentally observed
states are also presented here. Again we see from Table 6 that there
is quite a remarkable correspondence of the calculated KK excitations
for the $K\pi$-system with experimentally observed states of strange
mesons.

The Kaluza-Klein towers of KK excitations for the $K\pi\pi$-system
and $K\rho$-system are shown in Table 7 and Table 8 extracted from
\cite{10} where the comparison with experimentally observed mass
spectrum has been presented too. As in the previous history we see
from Tables 7--8 quite a remarkable correspondence of the calculated
KK excitations for the $K\pi\pi$-system and $K\rho$-system with the
masses of the states  where such states are experimentally observed.
Many blanks in Tables 7--8 indicate a wide field in experimental
study of the $K2\pi$-system.

We can see new recently observed states $D_{sJ}(2317)$ decaying to
$D_s\pi^0$ and $D_{sJ}(2457)$ decaying to $D^*_s\pi^0$ (discussed in
comprehensive review talk presented at this Conference by Belle
Collaboration \cite{13}) in Tables 9-10 extracted from Ref.
\cite{14}, where the Kaluza-Klein towers of KK excitations for the
$D_s\pi$-system and $D^*_s\pi$-system are presented. We would also
like to point out that $D_{sJ}(2317)$ state may occupy
$M_{22}^{K^*K}$-storey in the Kaluza-Klein tower of KK excitations
for the $K^*K$-system; see Table 11.  The most impressive fact from
the view point of our developed theoretical conception is the first
observation of a very narrow charmonium state with a mass of
$3871.8\pm0.7(stat)\pm0.4(syst)\,MeV$ which decays into
$\pi^+\pi^-J/\psi$ \cite{13,15}. It has been stressed \cite{15} that
the $\pi^+\pi^-$ invariant mass for the $M(3872)$ signal region
concentrate near the $\rho$ mass. Such state really exists, and it
lives just on the second storey in the Kaluza-Klein tower of KK
excitations for the $\rho J/\psi$-system; see Table 12. As is seen
from Table 12 there is a wonderful agreement of experimentally
measured mass with theoretically calculated one.

Really, a wealth of new and exciting experimental data have been
presented at the Conference HADRON'03. Here we would also like to
concern the recent results of E835 Experiment at Fermilab presented
in Aschaffenburg in talks  \cite{18} and  \cite{19}. It was pleased
for us to hear that E835 have precisely measured directly the mass
and width of $\eta_c(1{}^1S_0)$ in $\bar pp$ annihilation:
$M(\eta_c)=2984.1\pm 2.1\pm 1.0$ MeV and
$\Gamma(\eta_c)=20.4^{+7.7}_{-6.7}\pm 2.0$ MeV \cite{18}. This new
E835 measurement just filled the $M_{28}^{p\bar p }$-storey of the
Kaluza-Klein tower shown in Table~1.

New observations of $\bar pp \rightarrow \chi_0 \rightarrow
\pi^0\pi^0, \eta\eta$ through interference with the continuum and
precise measurements of mass and width ($M(\chi_0)=3415.5\pm 0.4\pm
0.07$ MeV and $\Gamma(\chi_0)=10.1\pm 1.0$ MeV) \cite{18}  are also
nice news for us. As is seen from Table 13  $\chi_0$-state just
occupied the $M_{39}^{\eta\eta}$-storey of Kaluza-Klein tower for the
$\eta\eta$ system \cite{20}.

In the same Table 13 new (preliminary though) results of E835
Collaboration \cite{19} for the masses of resonances decaying into
$\eta\eta$ have been shown by bold-face numbers. New results of E835
Collaboration \cite{19} for the masses of resonances decaying into
$\eta\pi$ have been presented in Table 14  by bold-face numbers too
\cite{20}. Asterisks in Tables 13-14 mark the states which have not
been seen before. It was a great pleasure to establish that new E835
Collaboration results provided an additional excellent confirmation
of our theoretical conception.

\section{Mein Ruf to search new states}

As was mentioned above, we have performed an analysis of experimental
data on mass spectrum of the states containing strange mesons and
compared them with the calculated values provided by Kaluza-Klein
scenario \cite{10}. By this way we have found out quite
an interesting correspondence shown below\\

\centerline{7-storey:}
\[
(?)\sigma(650)\in M_7^{\pi\pi}(640-644)\quad\longleftrightarrow \quad
K^*(892)\in M_7^{K\pi}(893-898),
\]

\centerline{15-storey:}
\[
f_2(1275)\in M_{15}^{\pi\pi}(1273-1275)\quad\longleftrightarrow \quad
K_2^*(1430)\in M_{15}^{K\pi}(1431-1434),
\]

\centerline{17-storey:}
\[
f_{0,2}(1430)\in M_{17}^{\pi\pi}(1435-1438)\quad\longleftrightarrow
\quad K_2^*(1580)\in M_{17}^{K\pi}(1579-1582),
\]

\centerline{18-storey:}
\[
f_{0}(1522)\in M_{18}^{\pi\pi}(1518-1520)\quad\longleftrightarrow
\quad K^*(1680)\in M_{18}^{K\pi}(1654-1657),
\]

\centerline{19-storey:}
\[
f_{0,?}(1580)\in M_{19}^{\pi\pi}(1599-1601)\quad\longleftrightarrow
\quad K_3^*(1780)\in M_{19}^{K\pi}(1730-1733),
\]

\centerline{22-storey:}
\[
\eta_{?,2}(1840)\in
M_{22}^{\pi\pi}(1845-1846)\quad\longleftrightarrow \quad
K_0^*(1950)\in M_{18}^{K\pi}(1960-1963),
\]

\centerline{23-storey:}
\[
f_{4}(1935)\in M_{23}^{\pi\pi}(1927-1928)\quad\longleftrightarrow
\quad K_4^*(2045)\in M_{23}^{K\pi}(2038-2040),
\]

\centerline{27-storey:}
\[
\rho_{5}(2250)\in M_{27}^{\pi\pi}(2256-2257)\quad\longleftrightarrow
\quad K_5^*(2380)\in M_{27}^{K\pi}(2352-2354).
\]\\
From this correspondence it follows that $K\pi$-system looks like a
system built from two-pion system by replacement of some one pion
with a kaon. In fact, all experimentally observed hadron states in
the $K\pi$-system have the corresponding partners in the two-pion
system. However, some hadron states in the two-pion system do not
have the corresponding strange partners in the $K\pi$-system
experimentally observed so far. That is why the further study of the
$K\pi$-system is quite a promising subject of the investigations.

Concerning the three-pion system we have found out that\\
\[
a_2(1311) \in M_{10}^{3\pi}(1309-1313),\qquad a_2(1311) \in
M_{10}^{\rho\pi}(1310-1312).
\]\\
Moreover, the strange partner of the $a_2$-meson is predicted which
we
would like to call as $a_2^s$-meson\\
\[
\fbox{$\displaystyle a_2^s(1520) \in M_{10}^{K2\pi}(1517-1523),\qquad
a_2^s(1520) \in M_{10}^{K\rho}(1519-1522)$}\,.
\]\\
Apart of isospin $a_2^s(1520)$-meson may have the same quantum
numbers as $a_2(1311)$-meson. We call up to search the
$a_2^s(1520)$-meson and other strange partners of the three-pion
states experimentally observed till now \cite{16}. In this respect it
seems the factory with intensive kaon beams would be a very good
device to realize such programm. However, we would like to especially
emphasize that recently observed states in the  $K_S^0K_S^0$ system
reported at this Conference by ZEUS Collaboration \cite{12} seem
indicate on the possibility to observe at HERA the
$a_2^s(1520)$-meson in the  $K_S^0\pi^+\pi^-$ system where the
invariant mass of the $\pi^+\pi^-$ system concentrated near the
$\rho$-peak.

\section{One comment}

In the consideration made above we have used the simplest form of
torus for the internal compact extra space and considered only
diagonal elements in the Kaluza-Klein towers. In fact, we have
established the non-trivial physical principle according to which KK
modes of decay products preferably paired up in compound system when
they lived on one and the same storey in Kaluza-Klein tower. However,
there are exceptional cases. For example, $\rho$ and $\omega$ mesons
appear as the non-diagonal elements of the Kaluza-Klein towers:
\begin{equation}\label{rho}
m_\rho\in M_{n,m}^{\pi^1\pi^2}= \sqrt{m_{\pi^1}^2+\frac{n^2}{R^2}} +
\sqrt{m_{\pi^2}^2+\frac{m^2}{R^2}},
\end{equation}
\[
M_{n,m}^{\pi^+\pi^-}(n_{\pi^+}=12,
m_{\pi^-}=4)=766.97\mbox{MeV},\quad
M_{n,m}^{\pi^+\pi^-}(n_{\pi^+}=13, m_{\pi^-}=4)=773.85\mbox{MeV},
\]
\[
M_{n,m}^{\pi^0\pi^0}(n=13, m=4)=769.78\mbox{MeV},\quad
M_{n,m}^{\pi^+\pi^0}(n_{\pi^+}=13, m_{\pi^0}=4)=770.92\mbox{MeV},
\]
and
\begin{equation}\label{omega}
m_\omega\in M_{n,m,k}^{\pi^+\pi^-\pi^0}=
\sqrt{m_{\pi^+}^2+\frac{n^2}{R^2}} +
\sqrt{m_{\pi^-}^2+\frac{m^2}{R^2}} +
\sqrt{m_{\pi^0}^2+\frac{k^2}{R^2}},
\end{equation}
\vspace{2mm}
\[
M_{n,m,k}^{\pi^+\pi^-\pi^0}(n_{\pi^+}=5,m_{\pi^-}=6,k_{\pi^0}=5)=782.80\mbox{MeV}.
\]
In general, as it follows from the observed hadron spectrum, the
non-diagonal elements of the Kaluza-Klein towers are physically
suppressed. Actually, the architecture of the hadron Kaluza-Klein
towns is unambiguously defined by the internal compact extra space
with its geometry and shapes, and we have to learn much more about
the geometry and shapes of the compact internal extra space. However,
one very important point in Kaluza-Klein picture is established now
in a reliable way: the size of the internal compact extra space
defines the global characteristics of the hadron spectra while the
masses of the constituents are the fundamental parameters of the
compound systems which the elements of the global structures being.
The knowledge of the true internal compact extra space is the
knowledge of the Everything that is the God. Our consideration made
above has shown that we found out a good approximation to the true
internal extra space. In our opinion, the global goal of the Natural
Philosophy and the fundamental particle\&nuclear physics as its part,
in the future, will be in that to perceive the true internal extra
space.

\section{Conclusion}

No doubt, the year 2003 will enter the history of particle physics as
a year of fundamental discoveries. A series of new mesons have been
discovered whose properties are in a strong disagreement with the
predictions of conventional QCD-inspired quark potential models.

What is a remarkable here is that all new narrow states have been
observed at the masses which are surprisingly far from the
predictions of conventional quark potential models. It is still more
remarkable that all new observed states are very narrow, their total
widths being about a few MeV. The small widths were found to be in
contradiction with quark model expectations as well. Does it mean the
end of the constituent quark model? In any case, this means either
considerable modifications in the conventional quark models have to
be introduced or that completely new approaches should be applied in
hadron spectroscopy.

We claim that existence of the extra dimensions in the spirit of
Kaluza and Klein together with some novel dynamical ideas may provide
new conceptual issues for the global solution of the spectral problem
in hadron physics.

In fact, we have shown that one simple formula with one fundamental
constant described more than 140 experimentally observed hadron
states. This is the most impressive fact in the theoretical
conception that has been developed. No quantum numbers were ascribed
to the predicted states because we have made only model independent
predictions for the masses of the states based on the existence of a
compact internal extra space. A special model under particular
consideration could give necessary information about quantum numbers
of the states and (super)fine splitting of the masses.

The performed analysis allows to conclude definitely that {\bf the
experimentally observed structures in hadron spectra reveal the
existence of extra dimensions and confirm the Kaluza-Klein picture of
the world}.

This is certainly a remarkable fact that a series of our publications
was followed by the fundamental discoveries in hadron spectroscopy
mentioned above, and here we would like to point out that strong time
correlation.

It is clear that further experimental studies with a higher mass
resolution are of great importance. In particular, this refers to the
problem of a large resonance overlap. It will also be important to
learn how one could experimentally extract fine structures in a broad
peak. We believe that the idea of an ultra-high resolution hadron
spectrometer \cite{6} is a vital experimental problem which can be
solved in hadron physics in the nearest future. Anyway, it is too
much desirable, and I do hope our most courageous wishes will come
true.

I started report with Max Planck's saying, and I would like to finish
it with the words of Max Planck as well.

\vspace{2mm}

{\it For it fell to this (atom) theory to discover, in the quantum
action, the long-sought key to the entrance gate into the wonderland
of spectroscopy, which since the discovery of spectral analysis had
obstinately defied all efforts to breach it.}


\rightline{Max Planck, Nobel Lecture, June 2, 1920}

\vspace{2mm}

Paraphrasing Max Planck I could say that the discovery of the
fundamental scale of the internal extra space with its geometry and
shapes provides the long-awaited key to the entrance gate into the
wonderland of hadron spectroscopy, which since the discovery of
strong forces had obstinately defied all efforts to open it.


\newpage
\vspace*{3cm}
\begin{center}
Table 1: Kaluza-Klein tower of KK excitations of $pp(p\bar p)$ system
and experimental data.

\vspace{5mm} \hbox to \hsize {\hss
\begin{tabular}{|c|c|l|l||c|c|l|l|}   \hline
n & $M_n^{pp}$\,MeV & $M_{exp}^{pp}$\,MeV & $M_{exp}^{p\bar p}$\,MeV
& n & $M_n^{pp}$\,MeV & $M_{exp}^{pp}$\,MeV & $M_{exp}^{p\bar
p}$\,MeV \\ \hline 1 & 1878.38 & 1877.5 $\pm$ 0.5 & 1873 $\pm$ 2.5 &
15 & 2251.68 & 2240 $\pm$ 5 & 2250 $\pm$ 15
\\ \hline
2 & 1883.87 & 1886 $\pm$ 1 & 1870 $\pm$ 10 & 16 & 2298.57 & 2282
$\pm$ 4 & 2300 $\pm$ 20  \\ \hline 3 & 1892.98 & 1898 $\pm$ 1 & 1897
$\pm$ 1 & 17 & 2347.45 & 2350 & 2340 $\pm$ 40
\\ \hline
4 & 1905.66 & 1904 $\pm$ 2 & 1910 $\pm$ 30  & 18 & 2398.21 &  & 2380
$\pm$ 10
\\ \hline
5 & 1921.84 & 1916 $\pm$ 2 & $\sim $ 1920 &  19 & 2450.73 &   & 2450
$\pm$ 10   \\ \cline{5-8}
  &         & 1926 $\pm$ 2 &   & 20 & 2504.90 &   & $\sim$
  2500    \\ \hline
  &         & 1937 $\pm$ 2 & 1939 $\pm$ 2 & 21 & 2560.61 &
  &  \\ \cline{5-8}
6 & 1941.44 & 1942 $\pm$ 2 & 1940 $\pm$ 1 & 22 & 2617.76 &  & $\sim$
2620
  \\ \cline{5-8}
  &         & $\sim$1945  & 1942 $\pm$ 5 & 23 & 2676.27 &  & \\ \hline
7 & 1964.35 & 1965 $\pm$ 2 & 1968 & 24 & 2736.04 & 2735 & 2710 $\pm$
20  \\ \cline{5-8}
  &         & 1969 $\pm$ 2 & 1960 $\pm$ 15 & 25 & 2796.99 &  &
  \\ \hline
8 & 1990.46 & 1980 $\pm$ 2 & $1990^{\,+15}_{\,-30}$ & 26 & 2859.05 &
& 2850 $\pm$ 5 \\ \cline{5-8}
  &         & 1999 $\pm$ 2 &  &  27 & 2922.15 &  &
\\ \hline
9 & 2019.63 & 2017 $\pm$ 3 & 2020 $\pm$ 3 & 28 & 2986.22 & & $\mathbf{2984 \pm 2.1 \pm 1.0}$  \\
\hline 10 & 2051.75 & 2046 $\pm$ 3 & 2040 $\pm$ 40  & 29 & 3051.20 &
&
   \\ \cline{5-8}
   &         & $\sim$2050 & 2060 $\pm$ 20 &  30 & 3117.04 &  &  \\ \hline
11 & 2086.68 & 2087 $\pm$ 3 & 2080 $\pm$ 10 &  31 & 3183.67 &  &  \\
\cline{5-8}
   &         &            & 2090 $\pm$ 20 & 32 & 3251.06 &  &  \\ \hline
   &         & $\sim$2122 & 2105 $\pm$ 15 & 33 & 3319.15 &  & \\ \cline{5-8}
12 & 2124.27 & 2121 $\pm$ 3 & 2110 $\pm$ 10 & 34 & 3387.90 &  & 3370
$\pm$ 10 \\ \cline{5-8}
  &         & 2129 $\pm$ 5 & 2140 $\pm$ 30 & 35 & 3457.28
&  & \\ \hline 13 & 2164.39 & $\sim$2150 & 2165 $\pm$ 45 & 36 &
3527.25 &   & $h_c(1P)(3526)$ \\ \cline{5-8}
   &         & 2172 $\pm$ 5 & 2180 $\pm$ 10 & 37 & 3597.77 &
   & 3600
$\pm$ 20         \\ \hline 14 & 2206.91 & 2192 $\pm$ 3 & 2207 $\pm$
13 & 38 & 3668.81 &  &  \\ \hline
\end{tabular}
\hss}
\end{center}
\newpage
\vspace*{2cm}
\begin{center}
Table 2: Kaluza-Klein tower of KK excitations for two-pion system\\
and experimental data.

\vspace{5mm}
\begin{tabular}{|c|c|c|c|c|}\hline
 n & $ M_n^{\pi^0\pi^0}MeV $ & $ M_n^{\pi^0\pi^\pm}MeV $ & $
 M_n^{\pi^\pm\pi^\pm}MeV $ & $ M_{exp}^{\pi\pi}\,MeV $  \\
 \hline
1  & 282.41  & 286.80  & 291.21   & $\sim$ 300  \\
2  & 316.87  & 320.80  & 324.73   & 322 $\pm$ 8  \\
3  & 367.18  & 370.58  & 373.98   & 370 -- i356   \\
4  & 427.78  & 430.71  & 433.64   & 430 -- i325   \\
5  & 494.92  & 497.45  & 499.99   & 506 $\pm$ 10  \\
6  & 566.26  & 568.48  & 570.70   & 585 $\pm$ 20  \\
7  & 640.41  & 642.38  & 644.34   & 650 -- i370   \\
8  & 716.50  & 718.26  & 720.01   & 732 -- i123   \\
9  & 793.96  & 795.55  & 797.13   & 780 $\pm$ 30  \\
10 & 872.44  & 873.88  & 875.33   & 870 -- i370   \\
11 & 951.68  & 953.00  & 954.32   & 955 $\pm$ 10  \\
12 & 1031.50 & 1032.72 & 1033.94  & 1015 $\pm$ 15 \\
13 & 1111.78 & 1112.92 & 1114.05  &   \\
14 & 1192.43 & 1193.49 & 1194.55  & 1165 $\pm$ 50   \\
15 & 1273.38 & 1274.37 & 1275.36  & 1275.4 $\pm$ 1.2  \\
16 & 1354.57 & 1355.50 & 1356.43  & 1359 $\pm$ 40  \\
17 & 1435.96 & 1436.84 & 1437.72  & 1434 $\pm$ 18  \\
18 & 1517.53 & 1518.36 & 1519.19  & 1522 $\pm$ 25  \\
19 & 1599.24 & 1600.02 & 1600.81  & 1593 $\pm$ $8\,^{+\,29}_{-\,47}$
\\
20 & 1681.07 & 1681.82 & 1682.57  & 1678 $\pm$ 12  \\
21 & 1763.00 & 1763.72 & 1764.43  & 1768 $\pm$ 21  \\
22 & 1845.03 & 1845.71 & 1846.40  & 1854 $\pm$ 20  \\
23 & 1927.14 & 1927.79 & 1928.45  & 1921 $\pm$ 8  \\
24 & 2009.32 & 2009.94 & 2010.57  & 2010 $\pm$ 60  \\
25 & 2091.56 & 2092.16 & 2092.76  & 2086 $\pm$ 15  \\
26 & 2173.85 & 2174.43 & 2175.01  & 2175 $\pm$ 20  \\
27 & 2256.19 & 2256.75 & 2257.31  & $\sim$ 2250  \\
28 & 2338.58 & 2339.12 & 2339.66  & $\sim$ 2330  \\
29 & 2421.01 & 2421.53 & 2422.05  & 2420 $\pm$ 30  \\
30 & 2503.47 & 2503.97 & 2504.48  & 2510 $\pm$ 30  \\ \hline
\end{tabular}
\end{center}

\newpage
\vspace*{2cm}
\begin{center}
Table 3: Kaluza-Klein tower of KK excitations for three-pion system\\
and experimental data.

\vspace{5mm}
\begin{tabular}{|c|c|c|c|c|c|}\hline
 n & $ M_n^{3\pi^0}MeV $ & $ M_n^{\pi^{\pm}2\pi^0}MeV $ & $
 M_n^{\pi^0 2\pi^\pm}MeV $ & $ M_n^{3\pi^\pm}MeV $ & $
 M_{exp}^{3\pi}\,MeV $  \\
 \hline
1  & 423.62  & 428.02  & 432.42  & 436.81  &  \\
2  & 475.30  & 479.23  & 483.17  & 487.10  &  \\
3  & 550.77  & 554.17  & 557.57  & 560.98  & $\eta(0^{-+})[547]$ \\
4  & 641.68  & 644.60  & 647.53  & 650.46  &  \\
5  & 742.38  & 744.91  & 747.44  & 749.98  &  \\
6  & 849.40  & 851.61  & 853.83  & 856.05  &  \\
7  & 960.62  & 962.58  & 964.55  & 966.51  & $\eta'(0^{-+})[958]$ \\
8  & 1074.75 & 1076.51 & 1078.26 & 1080.02 &  \\
9  & 1190.95 & 1192.53 & 1194.12 & 1195.70 & 1194 $\pm$ 14 \\
10 & 1308.66 & 1310.10 & 1311.55 & 1312.99 & 1311.3$\pm$1.6 \\
11 & 1427.51 & 1428.84 & 1430.16 & 1431.49 & 1419 $\pm$ 31 \\
12 & 1547.25 & 1548.47 & 1549.69 & 1550.91 &  \\
13 & 1667.68 & 1668.81 & 1669.94 & 1671.08 & 1667 $\pm$ 4 \\
14 & 1788.65 & 1789.71 & 1790.76 & 1791.82 & 1801 $\pm$ 13 \\
15 & 1910.07 & 1911.06 & 1912.05 & 1913.04 &               \\
16 & 2031.86 & 2032.79 & 2033.72 & 2034.65 & 2030 $\pm$ 50 \\
17 & 2153.95 & 2154.83 & 2155.70 & 2156.58 & 2090 $\pm$ 30 \\
18 & 2276.29 & 2277.12 & 2277.95 & 2278.78 &  \\
19 & 2398.85 & 2399.64 & 2400.43 & 2401.22 &  \\
20 & 2521.69 & 2522.35 & 2523.10 & 2523.85 &  \\
21 & 2644.50 & 2645.22 & 2645.93 & 2646.65 &  \\
22 & 2767.54 & 2768.23 & 2768.91 & 2769.59 &  \\
23 & 2890.71 & 2891.36 & 2892.02 & 2892.67 &  \\
24 & 3013.97 & 3014.60 & 3015.23 & 3015.86 &  \\
25 & 3137.33 & 3137.94 & 3138.54 & 3139.14 &  \\
26 & 3260.78 & 3261.36 & 3261.94 & 3262.52 &  \\
27 & 3384.29 & 3384.85 & 3385.41 & 3385.97 &  \\
28 & 3507.87 & 3508.41 & 3508.95 & 3509.49 &  \\
29 & 3631.51 & 3632.03 & 3632.55 & 3633.08 &  \\
30 & 3755.21 & 3755.71 & 3756.21 & 3756.72 &  \\ \hline
\end{tabular}
\end{center}

\newpage
\vspace*{2cm}
\begin{center}
Table 4: Kaluza-Klein tower of KK excitations for $\rho\pi$ system\\
and experimental data.

\vspace{5mm}
\begin{tabular}{|c|c|c|c|}\hline
 n & $ M_n^{\rho\pi^0}MeV $ & $ M_n^{\rho\pi^\pm}MeV $ &
 $M_{exp}^{\rho\pi}MeV$  \\
 \hline
1  & 911.62  & 916.02  &  \\
2  & 932.19  & 936.13  &  \\
3  & 962.89  & 966.29  &  \\
4  & 1000.88 & 1003.81 &  \\
5  & 1044.23 & 1046.76 &  \\
6  & 1091.69 & 1093.91 &  \\
7  & 1142.48 & 1144.45 &  \\
8  & 1196.07 & 1197.83 &  \\
9  & 1252.08 & 1253.67 &  \\
10 & 1310.23 & 1311.67 & $a_2(2^{++})$ \\
11 & 1370.28 & 1371.60 &  \\
12 & 1432.05 & 1433.27 &  \\
13 & 1495.37 & 1496.50 &  \\
14 & 1560.10 & 1561.16 &  \\
15 & 1626.12 & 1627.11 & $\pi_1(1^{-+})$ \\
16 & 1693.32 & 1694.25 &  \\
17 & 1761.58 & 1762.46 &  \\
18 & 1830.83 & 1831.66 &  \\
19 & 1900.98 & 1901.77 &  \\
20 & 1971.95 & 1972.70 &  \\
21 & 2043.67 & 2044.39 &  \\
22 & 2116.10 & 2116.78 &  \\
23 & 2189.16 & 2189.81 &  \\
24 & 2262.81 & 2263.44 &  \\
25 & 2337.00 & 2337.60 &  \\
26 & 2411.69 & 2412.27 &  \\
27 & 2486.85 & 2487.41 &  \\
28 & 2562.43 & 2562.97 &  \\
29 & 2638.41 & 2638.93 &  \\
30 & 2714.76 & 2715.27 &  \\ \hline
\end{tabular}
\end{center}

\newpage
\vspace*{1cm}
\begin{center}
Table 5: Kaluza-Klein tower of KK excitations for two-kaon system\\
and experimental data.\footnote{The states labelled by (*) in the
$K_SK_S$ system  have been reported at this Conference in the talk of
E.~Fadeeva \cite{11}. The states labelled by (**) in the same
$K_SK_S$ system  have been reported at this Conference in the talk of
M.~Barbi (ZEUS Collaboration at HERA) \cite{12}. Bold-face number in
23-storey has been taken from the talk of I.~Vorobiev presented at
EPSHEP2003, 17--23 July 2003, Aachen, Germany.}

\vspace{5mm}
\begin{tabular}{|c|c|c|c|c|}\hline
 n & $ M_n^{2K^0}MeV $ & $ M_n^{K^0 K^\pm}MeV $ & $
 M_n^{2K^\pm}MeV $ & $ M_{exp}^{2K}\,MeV $  \\
 \hline
1  & 998.80  & 994.81  & 990.83  &   \\
2  & 1009.08 & 1005.14 & 1001.20 &   \\
3  & 1025.99 & 1022.11 & 1018.24 & 1019.417$\pm$0.014 \\
4  & 1049.21 & 1045.42 & 1041.63 &    \\
5  & 1078.32 & 1074.64 & 1070.95 & $X(1070)^{\,(*)}$  \\
6  & 1112.87 & 1109.30 & 1105.73 &  \\
7  & 1152.37 & 1148.93 & 1145.48 &    \\
8  & 1196.33 & 1193.01 & 1189.69 &    \\
9  & 1244.27 & 1241.08 & 1237.89 &  \\
10 & 1295.76 & 1292.69 & 1289.63 & $\sim 1300^{\,(**)}$ \\
11 & 1350.38 & 1347.44 & 1344.50 &   \\
12 & 1407.77 & 1404.96 & 1402.14 &  \\
13 & 1467.62 & 1464.91 & 1462.21 &   \\
14 & 1529.62 & 1527.02 & 1524.43 & ${1537^{\,+\,9}_{\,-\,8}}^{\,(**)}$ \\
15 & 1593.53 & 1591.04 & 1588.55 &   \\
16 & 1659.13 & 1656.74 & 1654.34 & 1655 $\pm$ 17 \\
17 & 1726.22 & 1723.92 & 1721.62 & $1726 \pm 7^{\,(**)}$ \\
18 & 1794.64 & 1792.43 & 1790.22 &  \\
19 & 1864.24 & 1862.11 & 1859.99 & 1864.1 $\pm$ 1.0  \\
20 & 1934.89 & 1932.85 & 1930.80 &  \\
21 & 2006.49 & 2004.52 & 2002.54 & $X(2000)^{\,(*)}$ \\
22 & 2078.93 & 2077.03 & 2075.12 &   \\
23 & 2152.14 & 2150.30 & 2148.45 & $\mathbf{2150 \pm 30}$ \\
24 & 2226.02 & 2224.24 & 2222.46 & $\xi(2230)$  \\
25 & 2300.53 & 2298.81 & 2297.08 &   \\
26 & 2375.60 & 2373.93 & 2372.26 &   \\
27 & 2451.17 & 2449.56 & 2447.94 & \\
28 & 2527.21 & 2525.64 & 2524.08 & \\
29 & 2603.67 & 2602.15 & 2600.63 &   \\
30 & 2680.52 & 2679.04 & 2677.57 &  \\ \hline
\end{tabular}
\end{center}

\newpage
\vspace*{2cm}
\begin{center}
Table 6: Kaluza-Klein tower of KK excitations for $K\pi$ system\\ and
experimental data.

\vspace{5mm}
\begin{tabular}{|c|c|c|c|c|c|}\hline
 n & $ M_n^{K^0 \pi^0}$MeV & $ M_n^{K^0 \pi^{\pm}}$MeV & $
 M_n^{K^\pm \pi^0}$MeV & $ M_n^{K^\pm \pi^\pm}$MeV & $
 M_{exp}^{K \pi}$\,MeV  \\
 \hline
1  & 640.60  & 645.00  & 636.62  & 641.02  &  \\
2  & 662.97  & 666.91  & 659.03  & 662.97  &   \\
3  & 696.58  & 699.99  & 692.71  & 696.11  &  \\
4  & 738.50  & 741.42  & 734.71  & 737.63  &  \\
5  & 786.62  & 789.16  & 782.93  & 785.47  &  \\
6  & 839.57  & 841.79  & 836.00  & 838.22  &  \\
7  & 896.39  & 898.36  & 892.95  & 894.91  & $K^*(892)$ \\
8  & 956.42  & 958.17  & 953.10  & 954.85  &  \\
9  & 1019.12 & 1020.70 & 1015.93 & 1017.51 &  \\
10 & 1084.10 & 1085.54 & 1081.03 & 1082.48 &  \\
11 & 1151.03 & 1152.35 & 1148.09 & 1149.41 &  \\
12 & 1219.64 & 1220.86 & 1216.82 & 1218.04 &  \\
13 & 1289.70 & 1290.83 & 1287.00 & 1288.13 &  \\
14 & 1361.03 & 1362.08 & 1358.43 & 1359.48 &  \\
15 & 1433.45 & 1434.44 & 1430.97 & 1431.95 & $K_{0,2}^*(1430)$ \\
16 & 1506.85 & 1507.78 & 1504.46 & 1505.39 &  \\
17 & 1581.09 & 1581.97 & 1578.79 & 1579.67 & $K_2(1580)$ \\
18 & 1656.08 & 1656.91 & 1653.87 & 1654.70 & $K^*(1680)$ \\
19 & 1731.74 & 1732.53 & 1729.61 & 1730.40 & $K_3^*(1780)$ \\
20 & 1807.98 & 1808.73 & 1805.93 & 1806.68 &  \\
21 & 1884.75 & 1885.46 & 1882.77 & 1883.49 &  \\
22 & 1961.98 & 1962.67 & 1960.08 & 1960.76 & $K_0^*(1950)$ \\
23 & 2039.64 & 2040.29 & 2037.80 & 2038.45 & $K_4^*(2045)$ \\
24 & 2117.67 & 2118.30 & 2115.89 & 2116.52 &  \\
25 & 2196.04 & 2196.65 & 2194.32 & 2194.92 &  \\
26 & 2274.72 & 2275.30 & 2273.06 & 2273.64 &  \\
27 & 2353.68 & 2354.24 & 2352.07 & 2352.63 & $K_5^*(2380)$ \\
28 & 2432.90 & 2433.44 & 2431.33 & 2431.87 &  \\
29 & 2512.34 & 2512.86 & 2510.82 & 2511.34 &  \\
30 & 2592.00 & 2592.50 & 2590.52 & 2591.02 &  \\ \hline
\end{tabular}
\end{center}

\newpage
\vspace*{2cm}
\begin{center}
Table 7: Kaluza-Klein tower of KK excitations for $K\pi\pi$ system\\
and experimental data.

\vspace{5mm}
\begin{tabular}{|c|c|c|c|c|c|}\hline
 n & $ M_n^{K^0 2\pi^0}$MeV & $ M_n^{K^0 2\pi^{\pm}}$MeV &
 $M_n^{K^\pm 2\pi^0}$MeV & $ M_n^{K^\pm 2\pi^\pm}$MeV & $
 M_{exp}^{K 2\pi}$\,MeV  \\
 \hline
1  & 781.81  & 790.61  & 777.83  & 786.62  &  \\
2  & 821.41  & 829.27  & 817.47  & 825.33  &  \\
3  & 880.17  & 886.98  & 876.30  & 883.10  &  \\
4  & 952.39  & 958.24  & 948.60  & 954.45  &  \\
5  & 1034.08 & 1039.15 & 1030.39 & 1035.46 &  \\
6  & 1122.70 & 1127.14 & 1119.13 & 1123.57 &  \\
7  & 1216.60 & 1220.53 & 1213.15 & 1217.08 &  \\
8  & 1314.67 & 1318.18 & 1311.35 & 1314.86 &  \\
9  & 1416.10 & 1419.27 & 1412.91 & 1416.08 &  \\
10 & 1520.32 & 1523.21 & 1517.25 & 1520.14 &  \\
11 & 1626.87 & 1629.51 & 1623.93 & 1626.57 & 1629 $\pm$ 7 \\
12 & 1735.39 & 1737.83 & 1732.57 & 1735.01 & 1730 $\pm$ 20 \\
13 & 1845.59 & 1847.86 & 1842.89 & 1845.15 & $\sim$ 1840 \\
14 & 1957.24 & 1959.36 & 1954.65 & 1956.76 &  \\
15 & 2070.14 & 2072.12 & 2067.66 & 2069.63 &  \\
16 & 2184.13 & 2186.00 & 2181.74 & 2183.60 &  \\
17 & 2299.07 & 2300.83 & 2296.78 & 2298.53 &  \\
18 & 2414.85 & 2416.51 & 2412.64 & 2414.30 &  \\
19 & 2531.36 & 2532.93 & 2529.23 & 2530.81 &  \\
20 & 2648.51 & 2650.01 & 2646.46 & 2647.96 &  \\
21 & 2766.25 & 2767.68 & 2764.27 & 2765.70 &  \\
22 & 2884.50 & 2885.86 & 2882.59 & 2883.96 &  \\
23 & 3003.21 & 3004.51 & 3001.36 & 3002.67 &  \\
24 & 3122.33 & 3123.58 & 3120.55 & 3121.80 &  \\
25 & 3241.82 & 3243.03 & 3240.10 & 3241.30 &  \\
26 & 3361.65 & 3362.81 & 3359.98 & 3361.14 &  \\
27 & 3481.78 & 3482.90 & 3480.16 & 3481.28 &  \\
28 & 3602.19 & 3603.27 & 3600.62 & 3601.70 &  \\
29 & 3722.85 & 3723.89 & 3721.32 & 3722.37 &  \\
30 & 3843.73 & 3844.74 & 3842.25 & 3843.26 &  \\ \hline
\end{tabular}
\end{center}

\newpage
\vspace*{2cm}
\begin{center}
Table 8: Kaluza-Klein tower of KK excitations for $K\rho$ system\\
and experimental data.

\vspace{5mm}
\begin{tabular}{|c|c|c|c|}\hline
 n & $ M_n^{K^0\rho}$\,MeV & $ M_n^{K^\pm\rho}$\,MeV & $
 M_{exp}^{K\rho}$\,MeV   \\
 \hline
1  & 1269.82 & 1265.83 &  \\
2  & 1278.30 & 1274.36 & 1273 $\pm$ 7 \\
3  & 1292.29 & 1288.42 &  \\
4  & 1311.59 & 1307.81 &  \\
5  & 1335.93 & 1332.24 &  \\
6  & 1365.00 & 1361.43 &  \\
7  & 1398.46 & 1395.01 & 1402 $\pm$ 7 \\
8  & 1435.99 & 1432.67 & 1414 $\pm$ 15 \\
9  & 1477.24 & 1474.05 & $\sim$ 1460 \\
10 & 1521.89 & 1518.82 &  \\
11 & 1569.63 & 1566.69 &  \\
12 & 1620.19 & 1617.37 &  \\
13 & 1673.29 & 1670.58 &  \\
14 & 1728.70 & 1726.10 & 1717 $\pm$ 27 \\
15 & 1786.20 & 1783.71 & 1776 $\pm$ 7 \\
16 & 1845.59 & 1843.20 &  \\
17 & 1906.71 & 1904.41 &  \\
18 & 1969.39 & 1967.18 & 1973$\pm$8$\pm$25 \\
19 & 2033.48 & 2031.35 &  \\
20 & 2098.86 & 2096.81 &  \\
21 & 2165.42 & 2163.44 &  \\
22 & 2233.05 & 2231.14 &  \\
23 & 2301.66 & 2299.82 &  \\
24 & 2371.16 & 2369.38 &  \\
25 & 2441.49 & 2439.76 &  \\
26 & 2512.57 & 2510.90 &  \\
27 & 2584.34 & 2582.72 &  \\
28 & 2656.75 & 2655.18 &  \\
29 & 2729.75 & 2728.22 &  \\
30 & 2803.29 & 2801.81 &  \\ \hline
\end{tabular}
\end{center}

\newpage
\vspace*{2cm}
\begin{center}
Table 9: Kaluza-Klein tower of KK excitations for $D_s^\pm\pi$ system\\
and experimental data.

\vspace{5mm}
\begin{tabular}{|c|c|c|c|c|c|}\hline
 n & $ M_n^{D_s^\pm \pi^0}$MeV & $ M_n^{D_s^\pm \pi^{\pm}}$MeV & $
 M_{exp}^{D_s^\pm \pi}$\,MeV   \\
 \hline
1  & 2110.64 & 2115.04 & $D_s^{*\pm}$(2112) \\
2  & 2129.18 & 2133.11 &    \\
3  & 2156.52 & 2159.92 &    \\
4  & 2189.87 & 2192.80 &    \\
5  & 2227.35 & 2229.89 &    \\
6  & 2267.80 & 2270.02 &    \\
7  & 2310.50 & 2312.47 & $D_{sJ}$(2317)   \\
8  & 2355.02 & 2356.77 &    \\
9  & 2401.06 & 2402.65 &    \\
10 & 2448.44 & 2449.88 &    \\
11 & 2497.02 & 2498.34 &    \\
12 & 2546.70 & 2547.92 &    \\
13 & 2597.40 & 2598.53 &    \\
14 & 2649.07 & 2650.13 &    \\
15 & 2701.66 & 2702.65 &    \\
16 & 2755.14 & 2756.07 &    \\
17 & 2809.45 & 2810.33 &    \\
18 & 2864.58 & 2865.41 &    \\
19 & 2920.50 & 2921.29 &    \\
20 & 2977.17 & 2977.92 &    \\
21 & 3034.59 & 3035.30 &    \\
22 & 3092.72 & 3093.40 &    \\
23 & 3151.54 & 3152.19 &    \\
24 & 3211.03 & 3211.66 &    \\
25 & 3271.18 & 3271.78 &    \\
26 & 3331.95 & 3332.53 &    \\
27 & 3393.34 & 3393.90 &    \\
28 & 3455.33 & 3455.87 &    \\
29 & 3517.90 & 3518.42 &    \\
30 & 3581.02 & 3581.53 &    \\ \hline
\end{tabular}
\end{center}

\newpage
\vspace*{2cm}
\begin{center} Table 10: Kaluza-Klein tower of KK excitations for
$D_s^{*\pm}\pi$ system and $D_{sJ}$(2457)-meson.

\vspace{5mm}
\begin{tabular}{|c|c|c|c|c|c|}\hline
 n & $ M_n^{D_s^{*\pm} \pi^0}$MeV & $ M_n^{D_s^{*\pm} \pi^{\pm}}$MeV & $
 M_{exp}^{D_s^{*\pm} \pi}$\,MeV  \\
 \hline
1  & 2254.01 & 2258.41 &    \\
2  & 2272.46 & 2276.39 &    \\
3  & 2299.65 & 2303.05 &    \\
4  & 2332.80 & 2335.73 &    \\
5  & 2370.02 & 2372.55 &    \\
6  & 2410.14 & 2412.36 &    \\
7  & 2452.47 & 2454.43 & $D_{sJ}^+$(2457) \\
8  & 2496.56 & 2498.31 &    \\
9  & 2542.12 & 2543.70 &    \\
10 & 2588.96 & 2590.41 &    \\
11 & 2636.96 & 2638.28 &    \\
12 & 2686.01 & 2687.23 &    \\
13 & 2736.04 & 2737.17 &    \\
14 & 2786.99 & 2788.05 &    \\
15 & 2838.82 & 2839.81 &    \\
16 & 2891.50 & 2892.43 &    \\
17 & 2944.98 & 2945.86 &    \\
18 & 2999.24 & 3000.07 &    \\
19 & 3054.26 & 3055.05 &    \\
20 & 3110.01 & 3110.76 &    \\
21 & 3166.46 & 3167.18 &    \\
22 & 3223.61 & 3224.30 &    \\
23 & 3281.43 & 3282.08 &    \\
24 & 3339.90 & 3340.53 &    \\
25 & 3399.00 & 3399.61 &    \\
26 & 3458.72 & 3459.30 &    \\
27 & 3519.04 & 3519.60 &    \\
28 & 3579.95 & 3580.49 &    \\
29 & 3641.42 & 3641.94 &    \\
30 & 3703.44 & 3703.94 &    \\ \hline
\end{tabular}
\end{center}

\newpage
\vspace*{2cm}
\begin{center}
Table 11: Kaluza-Klein tower of KK excitations for $K^*K$ system and
$D_{sJ}$(2317)-meson.

\vspace{5mm}
\begin{tabular}{|c|c|c|c|c|c|}\hline
 n & $ M_n^{K^{*\pm} K^0}$\,MeV & $ M_n^{K^{*0} K^{\pm}}$\,MeV & $
 M_{exp}^{K^{*}K}$\,MeV  \\
 \hline
1  & 1392.02 & 1392.48 & $h_1(?1^{+-})(1386\pm 19)$ \\
2  & 1400.05 & 1400.53 &   \\
3  & 1413.30 & 1413.82 &   \\
4  & 1431.57 & 1432.15 & $f_1(0^+1^{++})(1433\pm 0.8)$ \\
5  & 1454.63 & 1455.27 &   \\
6  & 1482.18 & 1482.89 & $\eta(0^+0^{-+})(1475\pm 5)$ \\
7  & 1513.94 & 1514.71 & $f_1(0^+1^{++})(1512\pm 4)$ \\
8  & 1549.58 & 1550.42 &   \\
9  & 1588.80 & 1589.70 &   \\
10 & 1631.30 & 1632.27 & $\eta_2(0^+2^{-+})(1632\pm 14)$ \\
11 & 1676.82 & 1677.83 &   \\
12 & 1725.08 & 1726.14 &   \\
13 & 1775.85 & 1776.95 &   \\
14 & 1828.91 & 1830.04 &   \\
15 & 1884.06 & 1885.22 &   \\
16 & 1941.12 & 1942.29 &   \\
17 & 1999.92 & 2001.11 &   \\
18 & 2060.32 & 2061.51 &   \\
19 & 2122.17 & 2123.38 &   \\
20 & 2185.37 & 2186.57 &   \\
21 & 2249.79 & 2251.00 &   \\
22 & 2315.35 & 2316.55 & $D_{sJ}$(2317) (?) \\
23 & 2381.94 & 2383.14 &   \\
24 & 2449.49 & 2450.68 &   \\
25 & 2517.92 & 2519.10 &   \\
26 & 2587.17 & 2588.34 &   \\
27 & 2657.17 & 2658.33 &   \\
28 & 2727.87 & 2729.01 &   \\
29 & 2799.22 & 2800.35 &   \\
30 & 2871.17 & 2872.28 &   \\ \hline
\end{tabular}
\end{center}

\newpage
\vspace*{2cm}
\begin{center}
Table 12 Kaluza-Klein tower of KK excitations for $\rho J/\psi$
system and $D_{sJ}$(3872)-meson.

\vspace{5mm}
\begin{tabular}{|c|c|c|c|}\hline
 n & $ M_n^{\rho J/\psi}$ MeV & $M_{exp}^{\rho J/\psi}$ MeV  \\
 \hline
1  & 3867.57 &   \\
2  & 3871.74 & 3871.8$\pm$0.7$\pm$0.4  \\
3  & 3878.67 &   \\
4  & 3888.30 &   \\
5  & 3900.58 &   \\
6  & 3915.41 &   \\
7  & 3932.73 &   \\
8  & 3952.42 &   \\
9  & 3974.39 &   \\
10 & 3998.54 &   \\
11 & 4024.75 &   \\
12 & 4052.92 &   \\
13 & 4082.95 &   \\
14 & 4114.74 &   \\
15 & 4148.19 &   \\
16 & 4183.22 &   \\
17 & 4219.74 &   \\
18 & 4257.68 &   \\
19 & 4296.95 &   \\
20 & 4337.48 &   \\
21 & 4379.23 &   \\
22 & 4422.11 &   \\
23 & 4466.09 &   \\
24 & 4511.11 &   \\
25 & 4557.11 &   \\
26 & 4604.07 &   \\
27 & 4651.93 &   \\
28 & 4700.65 &   \\
29 & 4750.21 &   \\
30 & 4800.57 &   \\ \hline
\end{tabular}
\end{center}

\newpage

\vspace*{2cm}

\begin{center}
Table 13: Kaluza-Klein tower of KK excitations for $\eta\eta$
system\\ and experimental data.

\vspace{5mm}
\begin{tabular}{|c|c|c|c|c|c|}\hline
 n & $ M_n^{2\eta}$\,MeV &
 $M_{exp}^{2\eta}$\,MeV &  n & $ M_n^{2\eta}$\,MeV &
 $M_{exp}^{2\eta}$\,MeV \\
 \hline
1  & 1097.74 & & 33 & 2948.47 &  \\
2  & 1107.10 & & 34 & 3025.66 &  \\
3  & 1122.54 & & 35 & 3103.15 &  \\
4  & 1143.80 & & 36 & 3180.91 &  \\
5  & 1170.56 & & 37 & 3258.94 &  \\
6  & 1202.47 & & 38 & 3337.19 &  \\
7  & 1239.11 & & 39 & 3415.68 & $\mathbf{\chi_{0}(3415.5\pm 0.4\pm 0.07)}$ \\
8  & 1280.10 & $f_2(1275)$ & 40 & 3494.36 &  \\
9  & 1325.01 & $\mathbf{2^{++}(1330 \pm 2)}$ & 41 & 3573.25 &  \\
10 & 1373.47 & & 42 & 3652.31 &  \\
11 & 1425.12 & & 43 & 3731.54 &  \\
12 & 1479.62 & $\mathbf{2^{++}(1477 \pm 5)}$ & 44 & 3810.93 &  \\
13 & 1536.66 & $f'_2(1525)$ & 45 & 3890.47 &  \\
14 & 1595.99 & $\pi_1(1600)$ & 46 & 3970.15 &  \\
15 & 1657.34 & & 47 & 4049.96 &  \\
16 & 1720.51 & $\mathbf{0^{++}(1734 \pm 4)}$ & 48 & 4129.90 &  \\
17 & 1785.29 & & 49 & 4209.95 &  \\
18 & 1851.53 & & 50 & 4290.11 &  \\
19 & 1919.07 & & 51 & 4370.38 &  \\
20 & 1987.78 & $\mathbf{4^{++}(1986 \pm 5)}$ & 52 & 4450.75 &  \\
21 & 2057.54 & $\mathbf{2^{++}(\sim 2030)}$ & 53 & 4531.21 &  \\
22 & 2128.24 & $\mathbf{2^{++}(2138 \pm 4)}$ & 54 & 4611.76 &  \\
23 & 2199.80 & & 55 & 4692.39 &  \\
24 & 2272.14 & & 56 & 4773.10 &  \\
25 & 2345.18 & $\mathbf{4^{++}(2352 \pm 8)^{*}}$ & 57 & 4853.89 &  \\
26 & 2418.86 & & 58 & 4934.75 &  \\
27 & 2493.13 & $\mathbf{6^{++}(2484 \pm 14)}$ & 59 & 5015.68 &  \\
28 & 2567.93 & & 60 & 5096.68 &  \\
29 & 2643.21 & & 61 & 5177.73 &  \\
30 & 2718.94 & & 62 & 5258.85 &  \\ \hline
\end{tabular}
\end{center}

\newpage

\vspace*{2cm}

\begin{center}
Table 14: Kaluza-Klein tower of KK excitations for $\eta\pi$ system\\
and experimental data.

\vspace{5mm}
\begin{tabular}{|c|c|c|c|}\hline
 n & $M_n^{\eta\pi^0}$\,MeV & $M_n^{\eta\pi^\pm}$\,MeV & $M_{exp}^{\eta\pi}$\,MeV \\
 \hline
1  & 690.08  & 694.47  &  \\
2  & 711.99  & 715.92  &  \\
3  & 744.86  & 748.26  &  \\
4  & 785.79  & 788.72  &  \\
5  & 832.74  & 835.28  &  \\
6  & 884.37  & 886.58  &  \\
7  & 939.76  & 941.73  &  \\
8  & 998.30  & 1000.05 &  \\
9  & 1059.49 & 1061.07 &  \\
10 & 1122.97 & 1124.40 &  \\
11 & 1188.40 & 1189.72 &  \\
12 & 1255.56 & 1256.78 &  \\
13 & 1324.22 & 1325.36 & $\mathbf{2^{++}(1330 \pm 2)}$ \\
14 & 1394.21 & 1395.27 & $1^{-+}(1400 \pm 20)$ \\
15 & 1465.36 & 1466.35 & $0^{++}(1474 \pm 20)$ \\
16 & 1537.54 & 1538.47 &  \\
17 & 1610.63 & 1611.51 &  \\
18 & 1684.53 & 1685.36 & $\mathbf{0^{++}(\sim 1700)}$ \\
19 & 1759.15 & 1759.94 & $\mathbf{2^{++}(1740 \pm 7)}$ \\
20 & 1834.42 & 1835.17 &  \\
21 & 1910.27 & 1910.98 &  \\
22 & 1986.64 & 1987.32 & $\mathbf{4^{++}(1986 \pm 5)}$ \\
23 & 2063.47 & 2064.12 &  \\
24 & 2140.73 & 2141.36 &  \\
25 & 2218.37 & 2218.97 & $\mathbf{4^{++}(2226 \pm 6)^{*}}$ \\
26 & 2296.36 & 2296.94 &  \\
27 & 2374.66 & 2375.22 &  \\
28 & 2453.25 & 2453.79 &  \\
29 & 2532.11 & 2532.63 &  \\
30 & 2611.21 & 2611.71 &  \\ \hline
\end{tabular}
\end{center}


\begin{thebibliography}{**}
\bibitem{1}
A.A.~Arkhipov, arXiv:hep-ph/0211449 (2002); preprint IHEP 2002-44,
Protvino, 2002, available at
http://dbserv.ihep.su/\~{}pubs/prep2002/ps/2002-44.pdf
\bibitem{2}
A.A.~Arkhipov, arXiv:hep-ph/0208215 (2002); preprint IHEP 2002-43,
Protvino, 2002, available at
http://dbserv.ihep.su/\~{}pubs/prep2002/ps/2002-43.pdf
\bibitem{3}
A.A.~Arkhipov, arXiv:hep-ph/0302164 (2003).
\bibitem{4}
A.A.~Arkhipov, arXiv:hep-ph/0302213 (2003).
\bibitem{5}
J.~Yonnet, B.~Tatischeff et al., Phys. Rev. C{\bf 63}, 014001-1
(2000).
\bibitem{6}
B.C.~Maglich, Controversies about fine structures and variable widths
in meson spectra and feasibility of an ultra-high resolution hadron
spectrometer, to appear in Proceedings of HADRON'03.
\bibitem{7}
J.~Button et al., Phys. Rev. {\bf 126}, 1858 (1962).
\bibitem{8}
A.A.~Arkhipov, arXiv:hep-ph/0304014 (2003).
\bibitem{9}
J.~Kuhn, this Conference.
\bibitem{10}
A.A.~Arkhipov, arXiv:hep-ph/0305167 (2003).
\bibitem{11}
E.~Fadeeva, this Conference.
\bibitem{12}
M.~Barbi (ZEUS Collaboration), this Conference; arXiv:hep-ex/0308006
(2003).
\bibitem{13}
P.~Krokovny (Belle Collaboration), this Conference;
arXiv:hep-ex/0307052, arXiv:hep-ex/0308019 (2003).
\bibitem{14}
A.A.~Arkhipov, arXiv:hep-ph/0306237 (2003).
\bibitem{15}
K.~Abe et al. (Belle Collaboration), arXiv:hep-ex/0308029 (2003).
\bibitem{16}
A.A.~Arkhipov, arXiv:hep-ph/0308321 (2003).
\bibitem{17}
A.A.~Arkhipov, arXiv:hep-ph/0309002 (2003).
\bibitem{18}
C.~Patrignani (for E835 Collaboration at FNAL), Charmonium
spectroscopy in $\bar pp$ annihilations, to appear in Proceedings of
HADRON'03.
\bibitem{19}
I.~Uman (for E835 Collaboration at FNAL), Observation of resonances
in $\bar pp\rightarrow \eta\eta\pi^0$ at 5.2 GeV/c, to appear in
Proceedings of HADRON'03.
\bibitem{20}
A.A.~Arkhipov, arXiv:hep-ph/0311370 (2003); preprint IHEP 2003-36,
Protvino, 2003.

\end{thebibliography}
\end{document}